\documentclass[10pt,aps,pra,twocolumn]{revtex4-2}

\usepackage{mathptmx}
\usepackage[T1]{fontenc}

\usepackage{graphicx}
\usepackage{hyperref}
\hypersetup{colorlinks = true, linkcolor = [rgb]{0.19411,0.51882,0.667058}, urlcolor = [rgb]{0.125490,0.29542,0.1647058}, citecolor = [rgb]{0.75882,0.37411,0.14117}}
\usepackage{amsmath}
\usepackage{amsfonts,upgreek}
\usepackage{amssymb}
\usepackage{braket}

\begin{document}

\title{Large baseline quantum telescopes assisted by partially distinguishable photons}

\author{Subhrajit Modak}
\author{Pieter Kok}
\affiliation{Department of Physics \& Astronomy, the University of Sheffield, Hounsfield Road, Sheffield, S3 7RH, UK.}

\begin{abstract}
\noindent
Quantum entanglement can be used to extend the baseline of telescope arrays in order to increase the spatial resolution. In one proposal by Marchese and Kok [Phys.\ Rev.\ Lett.\ \textbf{130}, 160801 (2023)], 
identical single photons are shared between receivers, and interfere with a star photon. In this paper we consider two outstanding questions: i) what is the precise effect of the low photon occupancy of the mode associated with the starlight, and ii) what is the effect on the achievable resolution of imperfect indistinguishability (or partial distinguishability) between the ground and star photons. We find that the effect of distinguishability is relatively mild, but low photon occupancy of the optical mode of the starlight quickly deteriorates the sensitivity of the telescope for higher auxiliary photon numbers. 
\end{abstract}

\maketitle

\noindent
Large baseline imaging is a well-known technique for improving the resolution of telescopes \cite{dravins2016intensity, czupryniak2023optimal, brown2023interferometric, bojer2022quantitative}. It has recently been used to resolve the spatial features of a black hole in the radio frequency (RF) spectrum \cite{akiyama2019first}, with a baseline comparable to the diameter of the Earth. This was possible because antennas can track both the amplitude and phase of RF waves and reconstruct the wave propagation characteristics. In contrast, at optical frequencies the amplitude and phase information must be retrieved via interferometry, which severely limits the baseline of optical telescopes. Light collected in the telescopes must travel through light pipes or optical fibers, which practically limits the baseline due to construction constraints or photon losses, respectively.

\begin{figure}[b]
\centering
\includegraphics[width=\columnwidth]{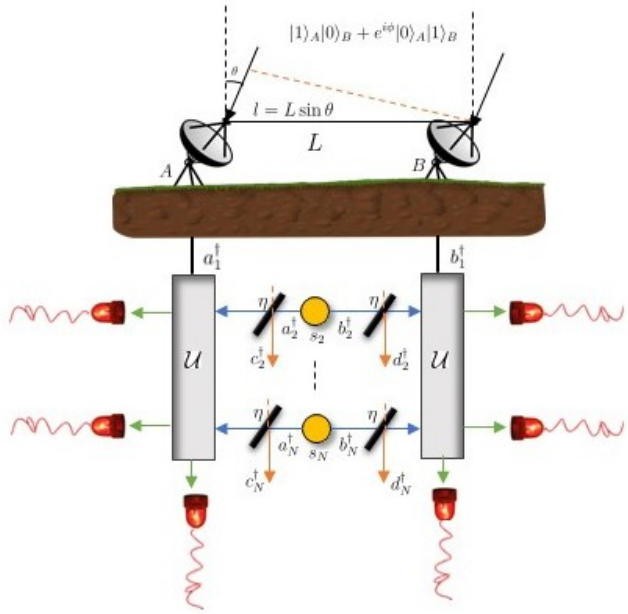}
\caption{ 
The proposal of Ref.~\cite{marchese2023large}: two receivers, $A$ and $B$, separated by a distance $L$, receive photons from a point-like astronomical object and from sources $S_2\ldots S_N$ located midway between the receivers. The path length of the star photon is kept short by co-locating the interferometers $U$ with the receivers. The photons sent from sources $S_2$ to $S_N$ experience reduced transmission $\eta^2 \leq 1$ due to fiber losses. The measurement statistics are used to estimate the declination angle $\theta$, encoding the position of the star.
}
\label{sar}
\end{figure}

Quantum technologies can be used to improve the performance of telescopes by choosing the optimal quantum observable \cite{kolobov2007quantum,tsang2016quantum,tsang2019resolving,pearce2017optimal,howard2019optimal,lupo2020quantum,zanforlin2022optical} or extending the baseline using quantum repeater protocols \cite{gottesman2012longer} and quantum error correction \cite{huang2022}. Recently, Marchese and Kok proposed a repeaterless method to extend the baseline of a telescope by employing multiple single-photon sources at the ground level that interfere at the receiver sites with photons originating from astronomical objects \cite{marchese2023large} (see Fig.~\ref{sar}). This minimizes the distance traveled by the photons that carry information about the astronomical objects and allows for redundancy in the ground-based photons to combat transmission losses over large distances. In this paper we address two outstanding open questions that arose from Ref.~\cite{marchese2023large}: i) what is the precise effect of the low photon occupancy of the mode associated with the starlight, and ii) what is the effect on the achievable resolution of imperfect indistinguishability (or partial distinguishability)
between the ground and star photons. 

This paper is organised as follows. In Sec.~\ref{sec:protocol} we recall the technical details of the repeaterless quantum telescope of Ref.~\cite{marchese2023large} and include the analysis of partially distinguishable photons. In Sec.~\ref{res} we establish the achievable resolution for the telescope, taking into account single mode photon occupancy. We conclude in Sec.~\ref{sec:conclusions}.

\section{The Quantum Telescope Protocol}\label{sec:protocol}\noindent
We assume that a point-like astronomical object, such as a distant star, is sufficiently far away that the incoming light can be effectively approximated by a plane wave. A photon from this object will arrive in two receivers of the telescope, $A$ and $B$ separated by a distance $L$, in the superposition state 
\begin{align}
 \ket{\psi_{\rm in}} = \frac{\ket{1}_A\ket{0}_B + e^{i\phi}\ket{0}_A\ket{1}_B}{\sqrt{2}}\, ,
\end{align}
where $\phi = kl = k L\sin\theta$ is the phase difference between the two receivers due to the path difference $l$ that originates from the direction $\theta$ of the incoming wave with wave number $k$ (see Fig.~\ref{sar}). The star photon is collected by the receivers $A$ and $B$, and interferes locally at the receivers with $N$ ground-based photons in interferometers $U$. The creation and annihilation operators of the ground-based modes are $\{a^\dagger_n,a_n\}$ and $\{b^\dagger_n,b_n\}$, and the input state becomes
\begin{equation}
\begin{split}
    |\psi\rangle_{\rm{tot}}^{\rm{in}}=&
    \left(\dfrac{1}{2}\right)^{\frac{N}{2}}\prod_{n=1}^N\left(a^\dagger_n+e^{i\phi \delta_{n,1}}b^\dagger_n\right)|0\rangle\, ,
    \label{eq:initialstateN}
\end{split}
\end{equation}
where the Kronecker delta ensures a the phase $\phi$ in mode 1, the starlight mode.
Here $|0\rangle$ is the vacuum state. The interferometers $U$ implement a discrete quantum Fourier transform (QFT), and detecting $D\leq N$ photons allows one to reconstruct the probability distributions $P_{\mathbf{d}}(\phi)$ in the presence of losses, where $\mathbf{d}$ indicates the detector signature. Ref.~\cite{marchese2023large} showed that this distribution contains a significant amount of information about the phase $\phi$, and therefore the angular position of the star $\theta$. We assume perfect number-resolving detectors.

\begin{figure}
    \centering
    \includegraphics[width=0.8\columnwidth]{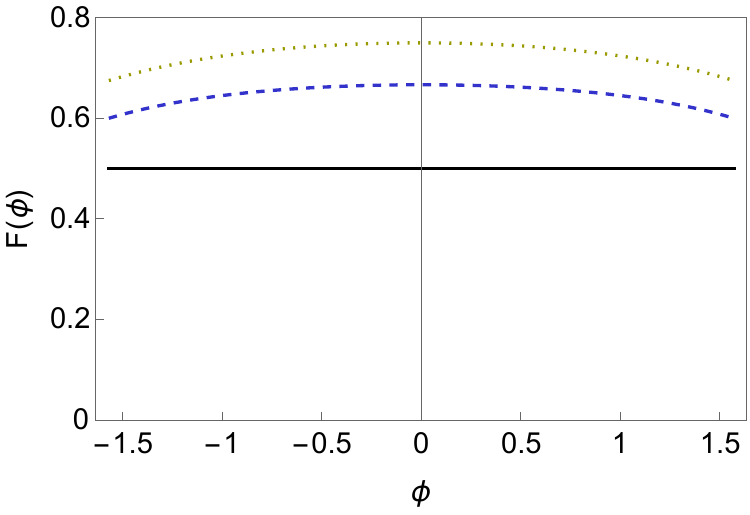}
    \caption{The dependence of the Fisher information on $\phi$ for various numbers $N$ of photons in the setup under the ideal condition of no loss (black: $N=2$, green: $N=3$, and red: $N=4$). The Fisher information increases with the photon number as $F_{N}(\phi)=1-1/N$.}
    \label{fig:linearityfisher}
\end{figure}

The Fisher information, which measures the information about $\phi$ in the probability distribution, is given by~\cite{marchese2023large}
\begin{equation}
     F_N(\phi)=\sum_{ \mathbf{d}}^{\sigma_N} P_{ \mathbf{d}}(\phi)\left(\dfrac{\partial \text{ln}P_{ \mathbf{d}}(\phi)}{\partial \phi}\right)^2\, ,
     \label{eq:FisherN}
\end{equation}
where $\sigma_N$ is the total number of possible detector outcomes. Including transmission losses for the ground photons, the Fisher information reduces to 
\begin{equation}
     F_N^{\rm{loss}}=\sum_{k=0}^{N-1} p^k (1-p)^{N-1-k}\binom{N-1}{k} F_{N-k}^{'}.
    \label{eq:fisherLOSS}
\end{equation}
where $F_{N-k}^{'}$ is the Fisher information for $D=N-k$ detected photons,  $k$ is the number of photons lost, and $p$ is the probability of a single photon loss. We assume that the loss on the star photon is negligible, and that the dominant loss is from the fiber transmission losses over long distances. In the case of no loss ($p=0$), the Fisher information still depends on $\phi$, as shown in Fig.~\ref{fig:linearityfisher} for $N\geq 3$. Hence the optimal arrangement is to include a variable phase shift in one receiver that balanced the interferometer in such a way that the receivers `point' to the source, i.e., $\phi\sim0$.

\section{Distinguishability of photons}\label{app:N2}\noindent
Next, we will consider how the distinguishability between photons affects the Fisher information. We take the star photon as the reference photon, and study how imperfect mode matching of the ground-based photons to the star photon affects the resolution. Our approach is as follows: the spatio-temporal characteristics of the star photon are labeled with index $\mu$. A ground-based photon in a single mode $a_j$ with a slightly different spatio-temporal character can then be described by a mode that is superposition of mode $a^\mu$ and an orthogonal mode $a^\nu_j$. These modes have corresponding mode operators that  satisfy the usual bosonic commutation relation: $[\hat{a}^{\mu}, \hat{a}^{\dagger\nu}]=[\hat{b}^{\mu}, \hat{b}^{\dagger\nu}]=\delta_{\mu\nu}$ and $[\hat{a}^{\mu}, \hat{a}^{\nu}]=[\hat{b}^{\mu}, \hat{b}^{\nu}]=0$. There are two possibilities: all the ground-based photons may be different from the star photon but identical to each other, or all the ground-based photons are also slightly distinguishable from each other. In the latter case, the part of the mode for photon $j$ that is orthogonal to the star photon mode must also be orthogonal to the other ground based photons, and we require an extra index $j$ on the orthogonal mode: $a^\nu \to a^{\nu}_{j}$. We model the distinguishability between the photons using the parameter $\mathcal{I}\in[0, 1]$:
\begin{equation}
\begin{aligned}
\label{cx}
    \hat{a}_1 & = \hat{a}^{\mu}_1 \\
     \hat{a}_j & = \sqrt{\mathcal{I}}\hat{a}^{\mu}_j+\sqrt{1-\mathcal{I}}\hat{a}^{\nu_j}_j\\
     \hat{b}_1 & = \hat{b}^{\mu}_1 \\
    \hat{b}_j & = \sqrt{\mathcal{I}}\hat{b}^{\mu}_j+\sqrt{1-\mathcal{I}}\hat{b}^{\nu_j}_j  \, .
\end{aligned}
\end{equation}
In this work we focus on the problem of phase estimation with multiple ground-based photons, where the degree of indistinguishability $\mathcal{I}$ is assumed to be known and uniform across the photons.

The initial state of the incoming star photon and the ground-based photons is given by
\begin{equation}
     |\psi\rangle_{\rm{tot}}^{\rm{in}}= \dfrac{1}{\sqrt{2}}\left(a^{\dagger}_1+e^{i\phi}b^{\dagger}_1\right)\otimes \prod_{j=2}^N \dfrac{a^{\dagger}_j+b^{\dagger}_j}{\sqrt{2}}|0\rangle\, .
     \label{eq:initialstate2modes}
\end{equation}
Following the transformation in Eq.~(\ref{cx}), the distinguishability between photons is then modeled by 
\begin{equation}
\begin{split}
    &a_j^\dagger =\sqrt{\mathcal{I}} a_j^{\dagger\mu}+\sqrt{1-\mathcal{I}}a_j^{\dagger\nu_j}\, \\
    &b_j^\dagger =\sqrt{\mathcal{I}} b_j^{\dagger\mu}+\sqrt{1-\mathcal{I}}b_j^{\dagger\nu_j}\, ,
\end{split}
\end{equation}
where $j\in[2, N]$ accounts only the ground-based photon sources. 

As in Ref.~\cite{marchese2023large}, we analyse the photon loss of ground-based fiber transmission using the well-known beam splitter model where the transmissivity $\eta$ is determined by the fiber loss $\eta=e^{-L/4L_{0}}$,  where each photon
travels over a length $L/2$, and $L_{0}$ is the attenuation length of the fiber. For simplicity we assume that $L_0$ is the same for all fibers. The transformed mode operators for the ground-based photons are thus written as
\begin{equation}
\begin{split}
    &a_j^{\dagger\kappa}=\eta a^{\dagger\kappa}_j+\sqrt{1-\eta^2}c_j^{\dagger\kappa}\\
    &b_j^{\dagger\kappa}=\eta b^{\dagger\kappa}_j+\sqrt{1-\eta^2}d^{\dagger\kappa}_j,
\end{split}
\end{equation}
where $\kappa\in\{\mu,\nu_j\}$ indicates the mode of the ground-based photons and $\{c^\dagger_j,c_j\}$ ($\{d^\dagger_j,d_j\}$) are the vacuum field operators on either side of the installation. 

At each site, the ground-based photon modes are mixed with the star photon modes. This is done by implementing QFT to the set of input photonic modes. This transformation appears as a simple balanced beam-splitter in the case with two photons, generating the following outputs for the modes on the left.
\begin{align}
    a_{1,\text{out}}^{\dagger\kappa} & =\dfrac{1}{\sqrt{2}}\left(- a_{1,\text{in}}^{\dagger\kappa}+ a_{2,\text{in}}^{\dagger\kappa}\right) \cr 
      a_{2,\text{out}}^{\dagger\kappa} & =\dfrac{1}{\sqrt{2}} \left(a_{1,\text{in}}^{\dagger\kappa}+ a_{2,\text{in}}^{\dagger\kappa}\right),
\end{align}
and analogously for the $b_i$ modes on the right.
Therefore, for $N=2$ the overall initial state in Eq.~\eqref{eq:initialstate2modes} becomes
\begin{align}
\begin{split}
  \vert\psi\rangle_{\rm{tot}}^{\rm{out}}  & =\frac{1}{2}\Bigg(\frac{(-a^{\dagger\mu}_1+a^{\dagger\mu}_2)}{\sqrt{2}}+e^{i\phi}\frac{(-b^{\dagger\mu}_1+b^{\dagger\mu}_2)}{\sqrt{2}}\Bigg)~\otimes~ \\
              &\qquad \Bigg(\alpha\Big(\eta~\frac{(a^{\dagger\mu}_1+a^{\dagger\mu}_2)}{\sqrt{2}}+\sqrt{1-\eta^{2}}~c^{\dagger\mu}_2\Big)+\\
              &\qquad  \beta\Big(\eta~\frac{(a^{\dagger\nu}_1+a^{\dagger\nu}_2)}{\sqrt{2}}+\sqrt{1-\eta^{2}}~c^{\dagger\nu}_2)+\\
              &\qquad  \alpha\Big(\eta~\frac{(b^{\dagger\mu}_1+b^{\dagger\mu}_2)}{\sqrt{2}}+\sqrt{1-\eta^{2}}~d^{\dagger\mu}_2\Big)+\\
              &\qquad  \beta\Big(\eta~\frac{(b^{\dagger\nu}_1+b^{\dagger\nu}_2)}{\sqrt{2}}+\sqrt{1-\eta^{2}}~d^{\dagger\nu}_2\Big)\Bigg)\vert0\rangle,
\end{split}
\end{align}
where $\alpha=\sqrt{\mathcal{I}}$ and $\beta=\sqrt{1-\mathcal{I}}$ are the amplitudes corresponding to the relative indistinguishability and the opposite, respectively. For $N=2$ we do not need to disambiguate $\nu$ with an index 2.

To find the Fisher information, we must calculate the probabilities
\begin{equation}
P_{\mathbf{d}}(\phi)=|\langle\mathbf{d}|\psi\rangle_{\rm{tot}}^{\rm{out}}|^2,
\end{equation}
for the detector signatures $\vert\mathbf{d}\rangle= \vert\mathbf{d}^{\mu},\mathbf{d}^{\nu}\rangle$, where $|\mathbf{d}\rangle=|d_1,d_2,d_3,d_4\rangle$ and $d_i\in\{1,2\}$ is the number of photons found in detector $i$. Evidently, two photons can be distributed among various modes and detectors in a number of ways. The relative phase shift will turn to a global phase when both the photons are picked up by the same detector or by separate detectors on the same side. Therefore, we can only obtain information about $\phi$ when two photons are detected in the same mode and on different sides of the detectors. Only the following configurations will give rise to probabilities that contribute to the Fisher information:
\begin{equation}
\begin{split}
   & P_{\vert a^{\dagger\mu}_2, b^{\dagger\mu}_1\rangle}(\phi)=P_{\vert a^{\dagger\mu}_1, b^{\dagger\mu}_2\rangle}(\phi)=\frac{\eta^{2}\mathcal{I}}{8}(1-\cos\phi),\\
    &P_{\vert a^{\dagger\mu}_1, b^{\dagger\mu}_1\rangle}(\phi)=P_{\vert a^{\dagger\mu}_2, b^{\dagger\mu}_2\rangle}(\phi)=\frac{\eta^{2}\mathcal{I}}{8}(1+\cos \phi).\\
\end{split}\label{eq:2modeProb}
\end{equation}
We point out that no information about the correlation is available with only one observed photon. Hence $F'_1$ is always zero and the total Fisher information is
\begin{equation}
    F_2^{\text{loss}}(\phi)=\dfrac{1}{2}(1-p)\mathcal{I},
\end{equation}
where $p=1-\eta^{2}$ is the probability of losing a single photon. 

The situation gets quite different when an extra ground-based photon is added. We further emphasize that each photon at the ground-based is identifiable from the reference to different degrees.  Upon considering all the detection events, the Fisher information yields
\begin{align}\label{pl}
 F_3^{\text{loss}}(\phi)  = & \frac{1}{2} (1-p)^2 \frac{6 (1+\cos \phi)}{5+4 \cos \phi} \mathcal{I} _2 \mathcal{I} _3 \cr
 & + \frac{1}{2} (1-p)^2 \left[(\mathcal{I}_2+\mathcal{I}_3)-2\mathcal{I} _2\mathcal{I} _3\right]\cr
     & +\frac{1}{2}p \left(1-p\right) \left(\mathcal{I} _2+\mathcal{I}_3\right),
\end{align}
where $\mathcal{I}_{j}$ measures the degree of indistinguishability for $j$th ground-based photons. The Fisher information in Eq.~\eqref{pl} consists of terms that arise when all the photons are detected and a single photon is lost. Events with two photon losses are not relevant, since they do not contribute to the Fisher information of the $N=3$ case. As the number of ground-based photons increases, it is expected to contribute more to the Fisher information. However, the relative distinguishability of photons appears to limit this advantage. An improvement over this can still be made if the photons are kept identical, i.e., $\mathcal{I}_2=\mathcal{I}_3$. Next, we will examine the extent to which the number of ground-based photons can be increased.

\begin{figure}
\centering
\includegraphics[width=\columnwidth]{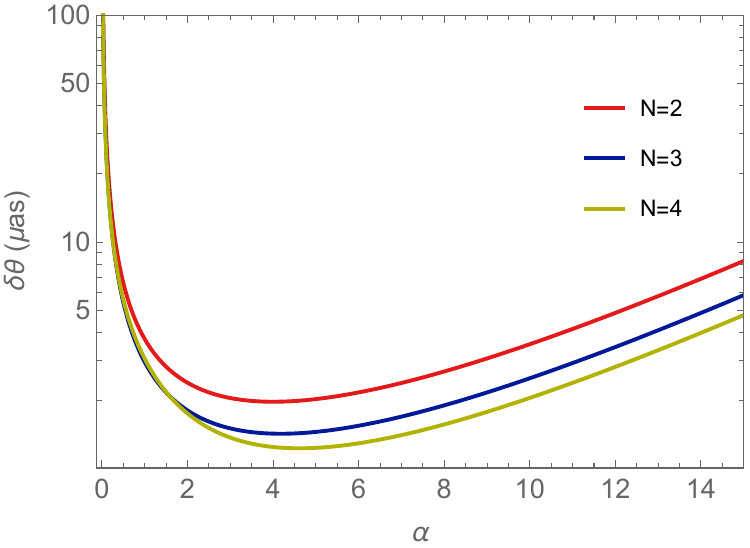}
\caption{The resolution angle $\delta\theta$ for identical photons at a given optical wavelength $\lambda$ = 628 nm and over a typical attenuation length $L_{0}$ = 10 km. In an ideal situation, we consider the star photon to arrive with certainty, i.e., $\epsilon=1$. The resolution angle decreases at short distances and approaches a minimum as the baseline increases. Further increase in baseline result in reduced resolution due to the predominant losses across the transmission channel. Thus, with more ground based photons we achieve better resolution and a shift in minimum is observed towards larger distances, allowing for an extension of the baseline.}
\label{obc}
\end{figure}

\begin{figure*}
\centering
{\includegraphics[width=0.32\textwidth]{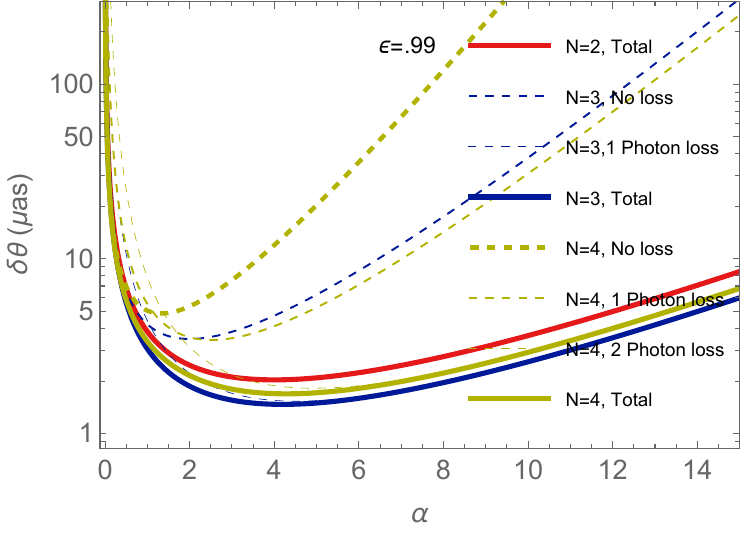}}\ 
{\includegraphics[width=0.32\textwidth]{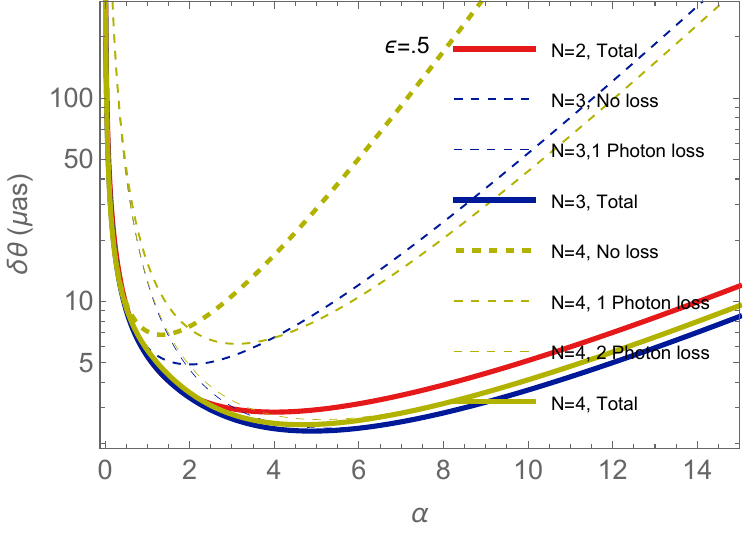}}\ 
{\includegraphics[width=0.32\textwidth]{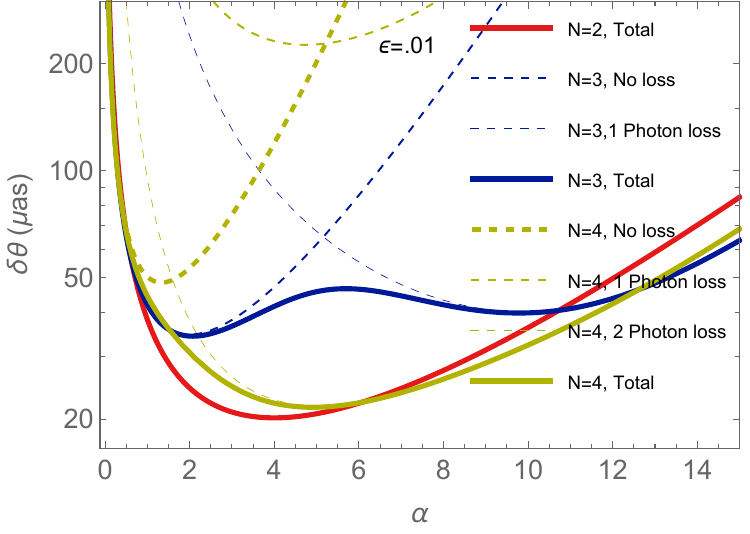}}
\caption{The resolution angle $\delta\theta$ as a function of baseline length $\alpha$ in units of attenuation length $L_{0}=10$~km. The curves are obtained using the same optical wavelength as in Fig.~\ref{obc} ($628$~nm) and different arrival probabilities of star photon. The ground-based photons are considered to be nearly identical (96\% indistinguishability) to the star photon. Figures from left to right indicate a decline in the rate of arriving star photon. Different solid colors shown in each figure correspond to a different total photon number. The resolution for the first two figures appears to improve as the number of ground-based photons increases up to $N=3$. However, because of complex interferences, increasing the number above $N = 3$ does not boost the resolution anymore. The improvement is reported to be broken even beyond $N=2$ when the arrival rate of star photon becomes highly uncertain, as shown in the picture to the right. Unlike the ideal case in Fig.~\ref{obc}, nearly identical ground-based photons do not always ensure lowering the minimum of resolution even when the arrival of star photon is slightly uncertain. However, until $N=3$ (Table ~\ref{adv}), the improvement is robust up to a moderate arrival rate of star photon.}
\label{sc}
\end{figure*}
\begin{figure*}
\centering
{\includegraphics[width=0.32\textwidth]{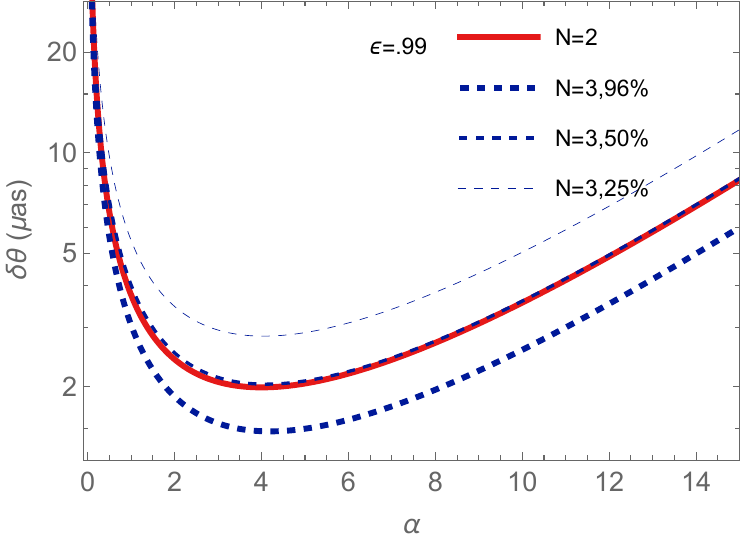}}\ 
{\includegraphics[width=0.32\textwidth]{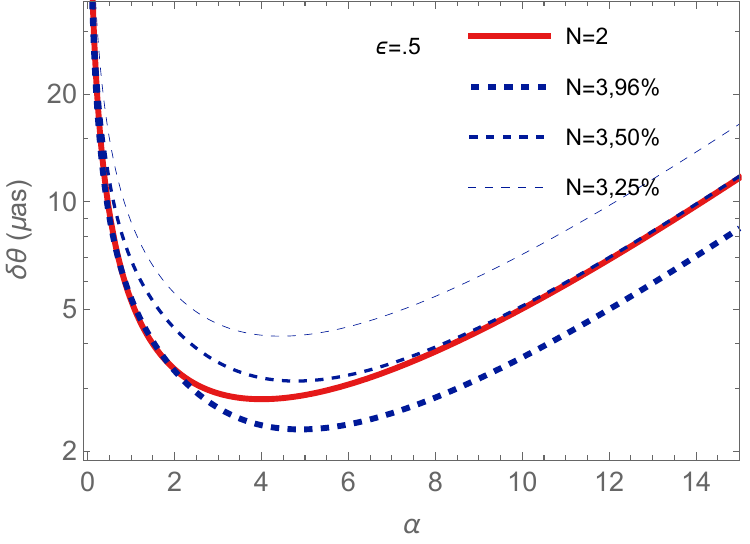}}\ 
{\includegraphics[width=0.32\textwidth]{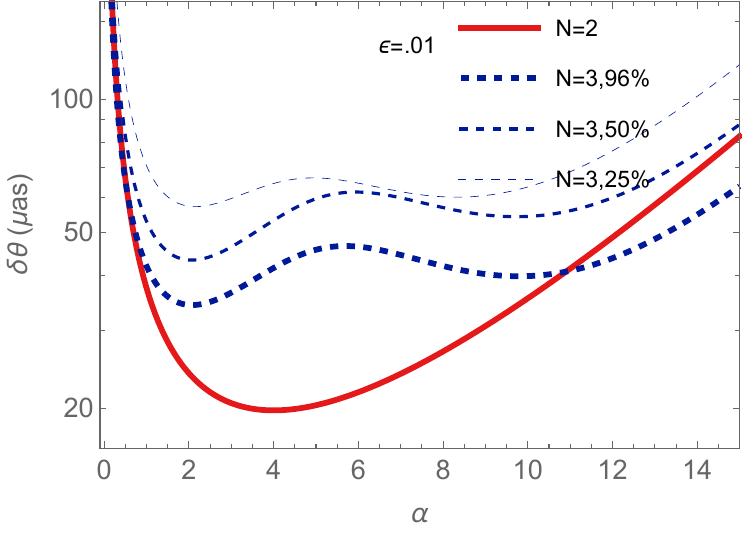}}
\caption{The resolution angle  $\delta\theta$ as a function of $\alpha$ for the same parametrization as in Fig.~\ref{sc}. The curves are also obtained for optical wavelengths $\lambda$ = 628~nm, with the same attenuation length scale $L_0 = 10$~km. The figures, from left to right, indicate different arrival probabilities of star photon. Each figure uses the solid red curve as a reference where the ground-based photon is indistinguishable from the stellar photon. The blue curves denote different levels of mutually distinguishable ground-based photons. The figure on the left shows that having more photons at the ground improves resolution, even if they can be distinguished to some extent. Once distinguishability reaches a particular threshold, the advantage is no longer available. The figure in the middle, when the arrival rate is much more uncertain than before, appears to follow the same pattern, but with a compromised resolution. Finally, if we take into account the situation where the rate of arrival is very uncertain, employing more ground-based photons is an unnecessary investment.  We rule out the case with $N=4$ since it appears to be an unsuitable option due to complex interference.}
\label{st}
\end{figure*}

\section{Achievable Resolution }\label{res}\noindent
Next, we consider the case of a weak thermal source at optical frequencies. The rate of photon emission $\epsilon$  within each coherence time interval is considerably less than one \cite{mandel1995optical}. Therefore, the density operator for the optical field in each interval becomes, to a good approximation, 
\begin{equation}
\rho = (1-\epsilon)\rho_0+\epsilon\rho_1,
\end{equation}
where $\rho_1$ is the one-photon state and $\rho_0=|0\rangle\langle 0|$ is the zero-photon state. Two-photon events are considered insignificant for the remainder of the discussion. The total probability of detecting $d$ photons in the $N$-photon setup becomes
\begin{equation}
    P_T(\mathbf{d})=(1-\epsilon) P_A(\mathbf{d})+\epsilon P_B(\mathbf{d}),
    \label{eq:probtot}
\end{equation}
where $P_A(\mathbf{d})$ and $P_B(\mathbf{d})$ are the probabilities for the absence and presence of the star photon, respectively. Clearly, we cannot identify from the detection of $d<N$ photons whether a star photon was present. In this case, the Fisher information becomes 
\begin{equation}
      F(\phi)=\sum_{\mathbf{d}}^{\sigma_D}\dfrac{\epsilon^2}{(1-\epsilon) P_A(\mathbf{d})+\epsilon P_B(\mathbf{d})}\left(\dfrac{\partial P_B(\mathbf{d})}{\partial \phi}\right)^2\, ,
\end{equation}
where $P_A(\mathbf{d})$ does not depend on $\phi$ because the absence of star photon carries no information about $\phi$. 

We calculate the Fisher information for various values of $N$. For $N=2$ the factor $\epsilon$, modifies the total Fisher information
\begin{equation}
    F_2(\phi)=\dfrac{1}{2}\epsilon(1-p)\mathcal{I}.
\end{equation}
Linear scaling in $\epsilon$ reflects the reduced rate of gaining information about $\phi$. For $N=3$, the scaling in  $\epsilon$ remains linear when all the photons are detected. However, the situation changes when a single photon is lost (see Appendix \ref{zz}). This implies a deteriorated estimation due to the uncertain arrival of star photon. Upon considering all the events, the total Fisher information results in
\begin{equation}\label{bb}
\begin{split}
    F_{3}(\phi)=   &\frac{\epsilon}{2} (1-p)^2  \frac{6 (1+\cos \phi)}{5+4 \cos \phi} \mathcal{I} _2 \mathcal{I} _3 \cr
 & + \frac{\epsilon}{2} (1-p)^2 \left[(\mathcal{I}_2+\mathcal{I}_3)-2\mathcal{I} _2\mathcal{I} _3\right] +\sum_{i} \mathcal{F}_{2}^{i},
\end{split}
\end{equation}
where $\mathcal{F}_{2}^{i}$ records various contributions when a single photon is lost, and $i$ runs over all such possible configurations (see Appendix \ref{zz}). In the limiting case, $\epsilon\rightarrow 1$ and $p\rightarrow 0$, the typical scaling with $N$ is reproduced by the Fisher information as $F_{N}\propto 1-1/N$. 

\begin{table}[t!]
\centering
\begin{tabular}{p{2cm} p{2cm} p{2cm} p{2cm}} 
 \hline  
 \centering
$\epsilon$ & $N$ & $\delta\theta_{\rm{min}} $($\mu\rm{as}$) & $\alpha_{\rm{opt}}$  \\[0.5ex]
\hline  
 \centering
  & 2 & 1.9797 & 4 \\ 
  \centering
 1 & 3 & 1.4311 & 4.1797 \\
 \centering
  & 4 & 1.2347 & 4.61273 \\[1ex] 
 \hline  
 \centering
  & 2 & 2.0303 & 4 \\ 
  \centering
 0.99 & 3 & 1.4698 & 4.19205 \\
 \centering
  & 4 & 1.6888 & 4.28221 \\[1ex] 
 \hline
  \centering
  & 2 & 2.8569 & 4 \\ 
  \centering
 0.5 & 3 & 2.3043 & 4.891 \\
 \centering
  & 4 & 2.4776 & 4.65838 \\[1ex] 
 \hline
  \centering
  & 2 & 20.2018 & 4 \\ 
  \centering
 0.01 & 3 & 34.2724 & 2.07328 \\
 \centering
  & 4 & 21.6339 & 4.90848 \\[1ex] 
 \hline
\end{tabular}

\caption{The table shows the parameters obtained with different $N$ for a given emission rate of star photon. The ground-based photons are considered to be nearly identical (96\%) to the star photon, and identical to each other. The third  column shows the minimum resolution $\delta\theta_{\rm{min}}$ at the optimal $\alpha_{\text{opt}}$, keeping the relative phase shift $\phi$ fixed close to zero.}
\label{adv}
\end{table}

Next, we look at the situation using four photons. We record an additional contribution to the Fisher information due to the higher number of photons. In order to make the expression analytically accessible, we assume all the photons are indistinguishable, while the case with distinct photons is calculated numerically.  This leads to
\begin{equation}\label{ee}
F_{4}(\phi)= 3 \epsilon  (1-p)^3\frac{ (9+7 \cos \phi)}{8 (5+3 \cos \phi)}+\sum_{i} \mathcal{F}_{3}^{i}+\sum_{j} \mathcal{F}_{2}^{j},
\end{equation}
where $\mathcal{F}_{3}^{i}$ records all the instances when a single photon is lost and $\mathcal{F}_{2}^{i}$ does the same when two photons are lost (see Appendix \ref{zz}).

The achievable resolution of the telescope is captured by the statistical error $\delta\theta$ in the angle $\theta = \phi/kL$, where we used the small angle approximation, since we are operating the telescope in the optimal regime where $\phi\approx 0$. From the error propagation formula we have  
\begin{equation}
    (\delta\theta)^2=\dfrac{(\delta\phi)^2}{k^2L^2},
\end{equation}
and $(\delta \phi)^2$ is lower bounded by the Fisher information via the Cram\'er-Rao bound. The best resolution is therefore given by
\begin{equation}
    (\delta\theta)^2=\left(\dfrac{1}{kL}\right)^2\dfrac{1}{F_N(\phi)}.
    \label{eq:varianceresolution}
\end{equation}
As expected, an increase in Fisher information leads to a better resolution. 

The resolution for identical photons and $\epsilon=1$ is shown in Fig.~\ref{obc} as a reference. Including low photon occupancy and ground-based photons that are 96\% identical to the star photon---and identical to each other---leads to the resolutions shown in Table~\ref{adv}, where we optimized the distance $L_{\rm opt}$ between the receivers. The last column includes $\alpha_{\rm opt} = L_{\rm opt}/L_0$. The values for $\delta\theta_{\rm min}$ are calculated for $L_0 = 10$~km and $\lambda = 628$~nm. Fig.~\ref{sc} shows the resolution as a function of $\alpha$. A minimum indicated the best possible resolution. We can see that the four-photon setup performs worse than $N=2$ and $N=3$ in most regimes.

In Table~\ref{table:results}, the resolution $\delta\theta$ is shown for the case where the ground-based photons are partially distinguishable from each other. The case of practical interest is how $N=3$ compares to $N=2$. For low occupancy $\epsilon=0.01$ the case $N=2$ is already optimal. Fig.~\ref{st} shows the resolution as a function of $\alpha$.

\begin{table}[t!]
\centering
\begin{tabular}{p{2cm} p{2cm} p{2cm} p{2cm}} 
 \hline  
 \centering
$\epsilon$ & $N$; $\mathcal{I}\%$ & $\delta\theta_{\rm{min}} $($\mu\rm{as}$) & $\alpha_{\rm{opt}}$  \\[0.5ex]
\hline  
 \centering
  & 2; 100 & 2.030 & 4 \\ 
  \centering
 0.99 & 3; 96 & 1.468 & 4.192 \\
 \centering
  & 3; 50 & 2.033 & 4.098 \\
  \centering
  & 3; 25 & 2.830 & 4.050 \\[1ex]
 \hline  
 \centering
  & 2; 100 & 3.999 & 4 \\ 
  \centering
  0.5 & 3; 96 & 2.3030 & 4.8911 \\
 \centering
  & 3; 50 & 3.1456 & 4.7868 \\
   \centering
  & 3; 25 & 4.2040 & 4.4409 \\[1ex]
 \hline
  \centering
   & 2; 100 & 19.980 
   & 4 \\ 
  \centering
  0.01 & 3; 96 & 34.272 & 2.0732 \\
 \centering
  & 3; 50 & 43.449 & 2.0621 \\
   \centering
  & 3; 25 & 57.230 & 2.1542 \\[1ex]
 \hline
\end{tabular}

\caption{The table presents the parameters for $N=3$ with varying degrees of distinguishability, using the identical $N=2$ as reference. The enhancement in resolution is observed with $N=3; \mathcal{I}=96\%$ for each case corresponding to a distinct emission rate. This advantage disappears as distinguishability increases even with an additional photon at the ground-based. As a result, boosting photon counts in practice does not necessarily guarantee improved estimation. }
\label{table:results}
\end{table}

\section{Conclusions}\label{sec:conclusions}\noindent
We considered the enhanced baseline two-receiver telescope, where quantum entanglement in the form of mode-entangled single photons are employed to help increase the baseline in the presence of loss. The astronomical photons carry the information about the object of interest, and minimizing their loss dictates that we measure them soon after they enter the telescope. The photons created on the ground do not carry information about the astronomical object, and we can therefore afford to lose some of them. Building redundancy by sending multiple photons to the receivers allows us to push the baseline much further, and as long as there is an appreciable photon flux in the detectors at the two receivers we can increase the practical resolution of the telescope. We measure the resolution using the mean square error on the angular position in the sky, and we calculate the classical Fisher information to find the resolution limit via the classical Cram\'er-Rao bound.

In this paper we considered two open questions, namely the effect of partial distinguishability of the photons, and the effect of low occupancy of thermal modes. When the occupancy $\epsilon=1$, the resolution is relatively robust against moderate photon distinguishability, and four nearly identical photons (96\% indistinguishability) strongly outperform the simple $N=2$ case for a resolution of $\delta\theta = 1.23~\upmu$as. However, even for a small reduction in $\epsilon=0.99$ and a indistinguishability of 96\% the four-photon setup is outperformed by the three-photon case. Hence, we conclude that in practice we do not need to consider more than three photons, including the astronomical photon. For an occupancy of $\epsilon=0.5$ and indistinguishability of 96\%, the three-photon case just outperforms the simplest case of $N=2$. However, the situation changes considerably when the occupancy is reduced further. At $\epsilon=0.01$, increased distinguishability of the photons has a pronounced detrimental effect on the $N=3$ case, but the leading cause of lower resolution is the reduced occupancy of the astronomical mode.

Another interesting effect is the contributions to the resolution of different numbers of measured photons in the $N=3$ case. There are two contributions to the resolution, namely when all photons are detected, and when only two out of three photons are detected. These two contributions have very different optimal baseline scales, and the resolution curve has two minima, shown in Fig.~\ref{sc}. As a result, for thermal light sources in the near infrared or optical frequency domain, the optimal setup is also the simplest setup, with one auxiliary photon that is filtered to be indistinguishable to the astronomical photon to within a few percent. Yet, even in this setup resolutions on the order of $\sim20~\upmu$as should be achievable.

\section*{Acknowledgements}\noindent
This research was funded by the Engineering and Physical Sciences Research Council through the grants Large Baseline Quantum-Enhanced Imaging Networks (Grant No. EP/V021303/1), and Mathematical Tools for Practical Quantum Imaging Protocols (Grant No. UKRI69). 

\providecommand{\noopsort}[1]{}\providecommand{\singleletter}[1]{#1}%
%


\widetext
\appendix\label{zz}
\section{Modified Fisher information}\noindent
We derive an expression for the modified Fisher information using two ground-based photons when a star emits photons at a rate $\epsilon$ substantially lower than one. The density operator for the initial state of the starlight is approximated as described in \ref{res}. We write the total input state with two possibilities, i.e., one in the absence (A) and other in presence (B) of a star photon:
\begin{equation} \label{eq1}
\begin{split}
\vert\psi\rangle_{A} & =\vert 0\rangle_{1}\otimes\vert1\rangle_{2}\otimes\vert 1\rangle_{3}, \\
\vert\psi\rangle_{B} & =\vert 1\rangle_{1}\otimes\vert1\rangle_{2}\otimes\vert 1\rangle_{3}
\end{split}
\end{equation}
Now, let the photons enter the interferometer and collect the state at the output:
\begin{align}
\vert1\rangle_{1}&\rightarrow\frac{1}{\sqrt{2}}\left(\frac{\left(e^{\frac{2 i \pi }{3}} a_{1}^{\dagger}+e^{-\frac{2 i \pi}{3}}  a_{2}^{\dagger}+ a_{3}^{\dagger}\right)}{\sqrt{3}}+\frac{e^{i\phi}\left(e^{\frac{2 i \pi }{3}}  b_{1}^{\dagger}+e^{-\frac{2 i \pi}{3}}  b_{2}^{\dagger}+ b_{3}^{\dagger}\right)}{\sqrt{3}}\right)_{1},\cr
\vert1\rangle_{2}&\rightarrow\frac{1}{\sqrt{2}}\left(\frac{\sqrt{1-p} \left(e^{-\frac{2 i \pi}{3}}  a_{1}^{\dagger}+e^{\frac{2 i \pi }{3}} a_{2}^{\dagger}+ a_{3}^{\dagger}\right)}{\sqrt{3}}+\sqrt{p} c_{2}^{\dagger}+\frac{\sqrt{1-p} \left(e^{-\frac{2 i \pi}{3}} b_{1}^{\dagger}+e^{\frac{2 i \pi }{3}}b_{2}^{\dagger}+b_{3}^{\dagger}\right)}{\sqrt{3}}+\sqrt{p} d_{2}^{\dagger}\right)_{2},\cr
\vert1\rangle_{3}&\rightarrow\frac{1}{\sqrt{2}} \left(\frac{\sqrt{1-p} ( a_{1}^{\dagger}+ a_{2}^{\dagger}+ a_{3}^{\dagger})}{\sqrt{3}}+\sqrt{p} c_{3}^{\dagger}+\frac{\sqrt{1-p} (b_{1}^{\dagger}+b_{2}^{\dagger}+b_{3}^{\dagger})}{\sqrt{3}}+\sqrt{p} d_{3}^{\dagger}\right)_{3}.
\end{align}
 We have treated all photons as identical and recorded every event associated with a single photon loss. The contributions to Eq.~\ref{bb} are as follows:
\begin{align}
\mathcal{F}_{2}^{1} & = \frac{ \epsilon ^2 p^2(1-p)^2  \sin ^2\phi }{3\left(\epsilon p(1-p) (1+\cos \phi)+4(1-\epsilon ) (1-p)^2\right)}, \nonumber \\
\mathcal{F}_{2}^{2}& =\frac{\epsilon ^2 p^2(1-p)^2 \left(\sin \phi +\sqrt{3} \cos \phi \right)^2}{3 \left(2\epsilon p (1-p)\left(2+\sqrt{3} \sin (\phi )-\cos (\phi )\right)+4  (1-\epsilon ) (1-p)^2\right)} ,\nonumber \\
\mathcal{F}_{2}^{3}& =\frac{ \epsilon ^2 p^2(1-p)^2 \left(\sin \phi -\sqrt{3} \cos \phi \right)^2}{3 \left(2 \epsilon p(1-p) \left(2-\sqrt{3} \sin \phi -\cos \phi \right)+4 (1-\epsilon )(1-p)^2 \right)}.
\end{align}
The situation where each ground-based photon has a distinct distinguishability from the reference photon is now examined. This will modify the contributions: 
\begin{align}
\mathcal{F}_{2}^{1} & =\frac{\alpha _2^4 \beta _3^4 \epsilon ^2 p^2(1-p)^2  \sin ^2 \phi }{6 \left(\epsilon  p(1-p)\alpha _2^2 \beta _3^2  (1+\cos \phi)+4  (1-\epsilon )(1-p)^2\alpha _2^2 \alpha _3^2\right)}, \nonumber \\
\mathcal{F}_{2}^{2} & =\frac{\alpha _3^4 \beta _2^4 \epsilon ^2 p^2(1-p)^2  \sin ^2 \phi }{6 \left(\alpha _3^2 \beta _2^2 \epsilon  p(1-p)  (1+\cos \phi)+4 \alpha _2^2 \alpha _3^2 (1-p)^2 (1-\epsilon )\right)},\nonumber \\
\mathcal{F}_{2}^{3}&=\frac{\alpha _3^4 \alpha _2^4\epsilon ^2 p^2(1-p)^2  \sin ^2 \phi }{3 \left(4 \alpha _2^2 \alpha _3^2 (1-p)^2 (1-\epsilon )+\alpha _2^2 \alpha _3^2 \epsilon  p(1-p)  (1+\cos \phi)\right)},\nonumber \\
\mathcal{F}_{2}^{4} &=\frac{\alpha _2^4 \beta _3^4 \epsilon ^2 p^2(1-p)^2  \left(\sin \phi-\sqrt{3} \cos \phi \right)^2}{6 \left(2 \alpha _2^2 \beta _3^2 \epsilon p (1-p)  \left(2-\sqrt{3} \sin \phi -\cos \phi\right)+4 \alpha _2^2 \alpha _3^2 (1-p)^2 (1-\epsilon )\right)},\nonumber \\
\mathcal{F}_{2}^{5} &=\frac{\alpha _3^4 \beta _2^4 \epsilon ^2 p^2(1-p)^2  \left(\sin \phi +\sqrt{3} \cos \phi \right)^2}{6 \left(2 \alpha _3^2 \beta _2^2 p(1-p) \epsilon  \left(2+\sqrt{3} \sin \phi -\cos \phi \right)+4 \alpha _2^2 \alpha _3^2 (1-p)^2(1-\epsilon )\right)},\nonumber 
\end{align}
\begin{align}
\mathcal{F}_{2}^{6} &=\frac{\alpha _2^4 \alpha _3^4\epsilon ^2 p^2(1-p)^2  \left(\sin \phi -\sqrt{3} \cos \phi \right)^2}{3 \left(4 \alpha _2^2 \alpha _3^2 (1-p)^2 (1-\epsilon )+2 \alpha _2^2 \alpha _3^2 p(1-p) \epsilon  \left(2-\sqrt{3} \sin \phi-\cos \phi\right)\right)},\nonumber \\
\mathcal{F}_{2}^{7}&=\frac{\alpha _3^4 \alpha _2^4\epsilon ^2 p^2(1-p)^2  \left(\sin (\phi )+\sqrt{3} \cos (\phi )\right)^2}{3 \left(4 \alpha _2^2 \alpha _3^2 (1-p)^2 (1-\epsilon )+2 \alpha _2^2 \alpha _3^2 p(1-p) \epsilon  \left(2+\sqrt{3} \sin \phi -\cos \phi\right)\right)},\nonumber \\
\mathcal{F}_{2}^{8}&=\frac{\alpha _2^4 \beta _3^4 \epsilon ^2 p^2(1-p)^2  \left(\sin \phi +\sqrt{3} \cos \phi \right)^2}{6 \left(2 \alpha _2^2 \beta _3^2 p (1-p) \epsilon  \left(2+\sqrt{3} \sin \phi -\cos \phi\right)+4 \alpha _2^2 \alpha _3^2 (1-p)^2 (1-\epsilon )\right)},\nonumber \\
\mathcal{F}_{2}^{9} &=\frac{\alpha _3^4 \beta _2^4 \epsilon ^2 p^2(1-p)^2  \left(\sin \phi-\sqrt{3} \cos (\phi )\right)^2}{6 \left(2 \alpha _3^2 \beta _2^2 p(1-p) \epsilon  \left(2-\sqrt{3} \sin \phi -\cos \phi\right)+4 \alpha _2^2 \alpha _3^2 (1-p)^2 (1-\epsilon )\right)},
\end{align} 
where $\alpha_j$ is the degree of distinguishability between the reference and $j^{th}$ ground-based photon. It is clear that the contribution from the loss terms begins scaling up linearly as $\epsilon$ is approaching 1. On the other hand, when $\epsilon$ is significantly lower than 1 refers to a situation where the rate of gaining information is reduced. Now, we carry out the same procedure using three ground-based photons. Unlike the earlier case, we have two sets of contributions: one for losing a single photon ($\mathcal{F}_{3}^{i}$) and other for losing two photons ($\mathcal{F}_{2}^{i}$). To keep things simple, we assume all the photons are identical whereas the cases with distinguishability has been studied numerically. The following are the contributions to Eq. \ref{ee} for a single photon loss:
\begin{align}
\mathcal{F}_{3}^{1} &=\frac{\epsilon ^2 p^2 (1-p)^4 \cos ^2\phi }{4 \left(2(1-\epsilon )(1-p)^3+\epsilon p (1-p)^2   (1-\sin \phi)\right)}\nonumber \\
\mathcal{F}_{3}^{2} &=\frac{\epsilon ^2 p^2 (1-p) ^4 \cos ^2\phi }{4 \left(2 (1-\epsilon )(1-p)^3 +\epsilon p(1-p)^2 (1+\sin \phi)\right)} \nonumber \\
\mathcal{F}_{3}^{3} &=\frac{6  \epsilon ^2 p^2 (1-p) ^4 \sin ^2(\phi )}{4 \left(18 (1-\epsilon)(1-p)^3+\epsilon p (1-p)^2   (5+4 \cos \phi)\right)} \nonumber \\
\mathcal{F}_{3}^{4} &=\frac{\epsilon ^2 p^2 (1-p)^4  \sin ^2\phi }{2 \left(2 (1-\epsilon)(1-p)^3 +\epsilon p (1-p)^2 (5-4 \cos \phi )\right)} \nonumber \\
\mathcal{F}_{3}^{5} &=\frac{\epsilon ^2 p^2 (1-p)^4 (\sin \phi +\cos \phi )^2}{2 \left(4(1-\epsilon ) (1-p)^3 +2\epsilon p (1-p)^2 (3+2 \sin \phi -2 \cos \phi)\right)} \nonumber \\
\mathcal{F}_{3}^{6} &=\frac{\epsilon ^2 p^2 (1-p)^4 (\sin \phi -\cos \phi )^2}{2 \left(4 (1-\epsilon )(1-p)^3+2\epsilon  p (1-p)^2  (3-2 \sin \phi -2 \cos \phi)\right)}
\end{align}
Similarly, the contributions are as follows when two photons are lost,
\begin{align}
\mathcal{F}_{2}^{1} &=\frac{p^4 (1-p)^2 \epsilon ^2 \sin ^2\phi }{8 \left(p^2 (1-p) \epsilon  (1+\cos \phi)+4 p (1-p)^2 (1-\epsilon )\right)}\nonumber \\
\mathcal{F}_{2}^{2} &=\frac{p^4 (1-p)^2 \epsilon ^2 \sin ^2\phi }{16 \left(p^2 (1-p) \epsilon  (1-\cos \phi )+2 p (1-p)^2 (1-\epsilon )\right)} \nonumber \\
\mathcal{F}_{2}^{3} &=\frac{p^4 (1-p)^2 \epsilon ^2 \cos ^2\phi }{16 \left(p^2 (1-p) \epsilon  (1+\sin \phi)+2 p (1-p)^2 (1-\epsilon )\right)} \nonumber \\
\mathcal{F}_{2}^{4} &=\frac{p^4 (1-p)^2 \epsilon ^2 \sin ^2\phi }{16 \left(p^2 (1-p) \epsilon  (1-\cos \phi )+2 p (1-p)^2 (1-\epsilon )\right)} \nonumber \\
\mathcal{F}_{2}^{5} &=\frac{p^4 (1-p)^2 \epsilon ^2 \cos ^2\phi }{16 \left(p^2 (1-p) \epsilon  (1-\sin \phi )+2 p (1-p)^2 (1-\epsilon )\right)}\nonumber \\
\mathcal{F}_{2}^{6} &=\frac{3 p^4 (1-p)^2 \epsilon ^2 \sin ^2\phi }{32 \left(p^2 (1-p) \epsilon  (1-\cos \phi )+4 p (1-p)^2 (1-\epsilon )\right)}\nonumber \\
\mathcal{F}_{2}^{7} &=\frac{3 p^4 (1-p)^2 \epsilon ^2 \sin ^2\phi }{32 \left(p^2 (1-p) \epsilon  (1+\cos \phi)+4 p (1-p)^2 (1-\epsilon )\right)}\nonumber 
\end{align}
\begin{align}
\mathcal{F}_{2}^{8} &=\frac{p^4 (1-p)^2 \epsilon ^2 \sin ^2\phi }{8 \left(p^2 (1-p) \epsilon  (\cos \phi +1)+4 p (1-p)^2 (1-\epsilon )\right)}\nonumber \\
\mathcal{F}_{2}^{9} &=\frac{p^4 (1-p)^2 \epsilon ^2 \sin ^2\phi }{16 \left(p^2 (1-p) \epsilon  (1-\cos \phi )+2 p (1-p)^2 (1-\epsilon )\right)}\nonumber \\
\mathcal{F}_{2}^{10}&=\frac{p^4 (1-p)^2 \epsilon ^2 \cos ^2\phi }{16 \left(p^2 (1-p) \epsilon  (\sin \phi +1)+2 p (1-p)^2 (1-\epsilon )\right)}\nonumber \\
\mathcal{F}_{2}^{11} &=\frac{p^4 (1-p)^2 \epsilon ^2 \sin ^2\phi }{16 \left(p^2 (1-p) \epsilon  (1-\cos \phi )+2 p (1-p)^2 (1-\epsilon )\right)}\nonumber \\
\mathcal{F}_{2}^{12} &=\frac{p^4 (1-p)^2 \epsilon ^2 \cos ^2\phi }{16 \left(p^2 (1-p) \epsilon  (1-\sin \phi )+2 p (1-p)^2 (1-\epsilon )\right)}\nonumber \\
\mathcal{F}_{2}^{13} &=\frac{p^4 (1-p)^2 \epsilon ^2 \sin ^2\phi }{32 \left(p^2 (1-p) \epsilon  (\cos \phi )+1)+4 p (1-p)^2 (1-\epsilon )\right)}\nonumber \\
\mathcal{F}_{2}^{14} &=\frac{p^4 (1-p)^2 \epsilon ^2 \sin ^2\phi }{32 \left(p^2 (1-p) \epsilon  (1-\cos \phi )+4 p (1-p)^2 (1-\epsilon )\right)}\nonumber \\
\mathcal{F}_{2}^{15} &=\frac{1}{2} (1-p) p^2 \epsilon
\end{align}
with $\epsilon\rightarrow 1$, both the set corresponding to the loss start scaling up linearly whereas for small $\epsilon$, the error in estimation becomes more dominant.

\end{document}